\documentclass{iau}
\usepackage[english]{babel}
\usepackage{graphicx}
\usepackage[utf8]{inputenc}
\usepackage{multicol}
\usepackage{amsmath, amssymb}
\usepackage{graphicx}
\usepackage{tabularx}
\usepackage{booktabs}
\usepackage{textcomp} 
\newcommand{\midsepremove}{\aboverulesep = 0mm \belowrulesep = 0mm}
\midsepremove
\newcommand{\midsepdefault}{\aboverulesep = 0.605mm \belowrulesep = 0.984mm}
\midsepdefault
\title[\small{Accelerating universe in modified teleparallel gravity theory}] 
{\small{Accelerating universe in modified teleparallel gravity theory}}
\author[\small{S.Shambel \etal} ]   
{\small{Shambel Sahlu$^{1,2}$, Joseph Ntahompagaze$^3$, Amare Abebe$^4$, and David F.Mota}$^5$}

\affiliation{$^{1}$ Astronomy and  Astrophysics Research and  Development Division, Entoto Observatory and Research Center, Ethiopian Space Science and Technology Institute, Addis Ababa, Ethiopia.\\ $^{2}$Department of Physics, College of Natural Science, Wolkite University, Wolkite, Ethiopia. $^3$Department of Physics, College of Science and Technology, University of Rwanda, Kigali, Rwanda.\\ $^4$Center for Space Research, North-West University, Mafikeng, South Africa.\\ $^5$ Institute of Theoretical Astrophysics, University of Oslo, Oslo, Norway}

\pubyear{2019}
\volume{356}  
\jname{Nuclear Activity in Galaxies Across Cosmic Time}
\editors{M. Povi\'c, P. Marziani, J. Masegosa, H. Netzer,\\ S. H. Negu, \&
	S. B. Tessema, eds.}

\begin{document}
\maketitle
\textbf{Abstract.} This paper studies the cosmology of accelerating expansion of the universe in 
 modified teleparallel gravity theory. We discuss the cosmology of $f(T,B)$ gravity theory and its implication
to the new general form of the equation of state parameter $w_{TB}$ for explaining the late-time accelerating expansion of the universe without the need for the 
cosmological constant scenario. We examine the numerical value of $w_{TB}$ in different paradigmatic  $f(T,B)$ gravity models. In those models, the numerical result of $w_{TB}$ is favored with observations in the presence of the torsion scalar $T$ associated with a boundary term $B$ and shows the accelerating expansion of the universe.

\section{Introduction}
To explain the cause behind the current cosmic accelerating expansion, different suggestions have been put forward. For instance, one suggestion  is that the cosmological constant is the one responsible  for this cosmic acceleration,  as presented in \cite{perlmutter2003supernova} and the second approach is the modification of General Relativity (GR) as \cite{clifton2012modified}.
In the second approach, several extra degrees of freedom are presented through the modification of GR to 
account for the present cosmic accelerating expansion and to study if the cosmic history from the early universe can produce this cosmic acceleration. 
This paper discusses how  the modified teleparallel gravity so-called,  $f(T,B)$ gravity scenario, is taken as an alternative approach for the $\Lambda$CDM model to describe the late-time accelerating expansion 
of the universe. We obtain the new expression of the equation of state parameter $w_{TB}$ in the effective torsion fluid.
\section{The cosmology of $f(T,B)$ gravity}
We consider different paradigmatic $f(T,B)$ gravity models, and in each model, we describe the accelerating universe in the late time by  plotting the  $w_{TB}$ versus redshift $z$. It is close to the well-known equation of state parameter $w =-1$. We start by providing the action that contains the $f(T,B)$ Lagrangian:
\begin{equation}\label{actionfTB}
 I_{f(T,B)} = \frac{1}{2\kappa^2}\int{d^4x e\Big[T+f(T,B)+L_{m}\Big]}\;,
 \end{equation}
 where $e$ is the determinate of the tetrad field $e^\mu_{A}$, $L_m$ is the matter Lagrangian and $\kappa^2= 8\pi G/c^4$ is the coupling constant\footnote{We assume that $\kappa^2= 8\pi G/c^4$ = 1.}.
 We assume that the total cosmic medium is composed of matter $\rho_m$, radiation $\rho_r$ and effective torsion fluid $\rho_{TB}$, and that one can directly derive the corresponding thermodynamic quantities in $f(T,B)$ gravity from eq. \eqref{actionfTB}, such as 
the energy density $\rho_{TB}$ and pressure  $p_{TB}$ of the torsion-like fluid as  presented in \cite{bahamonde2017noether}:
\begin{eqnarray}
 &&\kappa^2 \rho_{m} = -3H^2(3f_B+2f_B)+3H\dot{f}_B -3\dot{H}f_B +\frac{f(T,B)}{2} \;,\label{eq1} \\
 && \kappa^2 p_{m} = 3H^2+\dot{H})(3f_B+2f_T) 2H\dot{f}_T -\ddot{f}_B-\frac{f(T,B)}{2}\;. \label{eq2}
\end{eqnarray}
Therefore, the new general form of $w_{TB}$  can be constructed from the above equations for the effective torsion fluid defined as $w_{TB}=p_{TB}/\rho_{TB}$ and
is given by: 
\begin{eqnarray} \label{391}
  w_{TB}= -1+\frac{\ddot{f}_B -3H\dot{f}_B - 2\dot{H}f_T -2H\dot{f}_T}{3H^2(3f_B +2f_T)-3H\dot{f}_B +3\dot{H}f_B -{f(T,B)}/{2}}\; .
 \end{eqnarray}
 In the following we consider the well known two models namely: exponential and power-law $f(T)$ gravity models associated with the boundary term as
  $f(T,B) = B +f(T)$, for $B = 0$ it reads $f(T,B) = f(T)$ as presented in \cite{bahamonde2017noether2}.
  \section{Conclusions}
\begin{figure}[h!]
 \begin{minipage}{0.5\textwidth}
\includegraphics[width=0.75\textwidth]{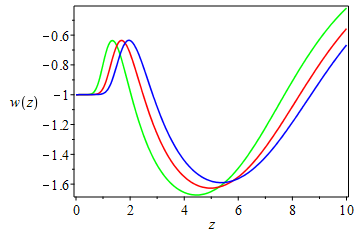}
    \caption{ $w(z)$ versus redshift $z$ for $f(T,B) =  B- T +\alpha T_0\left (1-e^{b\frac{T}{T_0}}\right) $ gravity  model  with different value of $b$. We use $b = 0.002$, $b = 0.004$ and $b = 0.006$ for  green, red and blue lines, respectively.}
    \label{fig:dustc121}
\end{minipage} 
\qquad
\begin{minipage}{0.5\textwidth}
\includegraphics[width=0.75\textwidth]{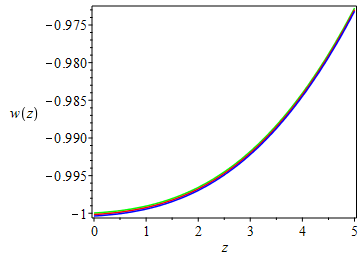}
    \caption{$w(z)$  versus redshift $z$ for  power-law $f(T,B) = B - T +\alpha(-T)^b$ gravity  model with different value of $b$. We use $b = 0.1$ for green line, $b = 0.4$ for red line and $b = 0.7$ for blue line.}
    \label{fig:dustc123}
 \end{minipage} 
 \end{figure}

Generally, in all models that are treated above, we numerically computed the effective equation of  state parameter
form of $w(z)$ through the use of eq. \eqref{391} and the value of $w(z)$  is favored with the observed value of the effective equation of  state parameter of cosmological constant $w \approx -1$ in the present universe.
So, we clearly show that all two  $f(T,B)$ gravity models can be regarded as an alternative way of cosmological constant model to describe the late-time accelerating expansion of the universe. 
Surprisingly, in all models the value of $w(z)$  in the present and near past universe 
asymptotically approaches to the equation of state parameter of the cosmological constant $w=-1$. For instance in Fig. \ref{fig:dustc121}, we clearly observe the history of the universe phase with the phantom-like $w(z)<-1$ and quintessence-like $w(z)>-1$ phases, while the other two models in Fig. \ref{fig:dustc123} show only the quintessence-like phase.  GR can be recovered for $B = b = 0$ for all models.  
  
\end{document}